\documentstyle[12pt]{article}
\nopagebreak
\newcommand\etal{{{\it et al.}\/}}

\def\oneh{{\textstyle {1\over 2}}}

\def\simle{{\ \lower2pt\hbox{$\sim $}\mkern-13mu \raise2pt \hbox{$<$}\ }}

\def\eq{\begin{equation}}
\def\ee{\end{equation}}

\def\eqa{\begin{eqnarray}}
\def\eea{\end{eqnarray}}

\def\eep{(e,e$'$p)\ }

\begin{document}


\centerline{\Large{\bf Multi-scattering effects in (e,e$'$p) knockout}}

\vskip 1.5cm

\centerline{\large{P.~Demetriou, S.~Boffi, C.~Giusti and F.~D.~Pacati}}

\vskip 1.0cm

\centerline{\small Dipartimento di Fisica Nucleare e Teorica, Universit\`a
di Pavia, and}

\centerline{\small Istituto Nazionale di Fisica Nucleare, 
Sezione di Pavia, Pavia, Italy}

\vskip 1.5cm


\begin{abstract}

\noindent 
Multistep direct scattering of the ejectile proton is considered in \eep
reactions in the quasi-free region as an improvement over the usual treatment of
final-state interactions by means of an optical potential. The theory is applied
to $^{40}$Ca\eep as a case example. Important contributions of two- and
three-step processes are found at high missing energy and momentum in agreement
with the experimental trend.

\end{abstract}

\bigskip



\section{Introduction}

Nucleon knockout by a high-energy electron in the quasi-free (QF) regime has
extensively been used as a powerful tool to investigate single-particle properties
of nuclei. The reason is that the electromagnetic interaction with nucleons is well
known from quantum electrodynamics and, in the one-photon-exchange approximation
and neglecting final-state interactions, the coincidence cross section for a
detected electron of energy $E_{k'}$ and angle $\Omega_{k'}$ and detected nucleon
of angle $\Omega$ has a factorised form:
\eq
{{\rm d}^3\sigma\over{\rm d}\Omega_{k'}{\rm d}E_{k'}{\rm d}\Omega} 
= K\sigma_{\rm ep}S({\vec p}_m,E_m),
\label{eq:factorised} 
\ee
where $K$ is a kinematical factor, $\sigma_{\rm ep}$ the elementary (off-shell)
electron-nucleon cross section and $S({\vec p}_m,E_m)$ the hole spectral function
depending on the missing momentum ${\vec p}_m$ and energy $E_m$. The momentum
${\vec p}_m$ is also the recoil momentum of the residual nucleus and $E_m$ its
excitation energy with respect to the target nucleus. Excitation spectra and 
momentum distributions of the produced hole have been measured for a variety of
nuclei along the whole periodic table (for a recent review, see
ref.~\cite{book96}). 

However, to extract from the data precise information, such as e.g. the values of
spectroscopic factors, an accurate treatment of final-state interactions (FSI) is
necessary with the result that the simple factorisation (\ref{eq:factorised}) is
destroyed~\cite{frullo79}. In the one-photon-exchange approximation the general
expression of the coincidence unpolarized cross section can be written in
terms of four structure functions $W_i$ as~\cite{book96}
\eqa
{{\rm d}^3\sigma\over {\rm d}\Omega_{k'}{\rm d}E_{k'}{\rm d}\Omega}
&=& { 2\pi^2\alpha \over \vert{\vec q}\vert}\, \Gamma_{\rm V}\, 
K  \{ W_{\rm T} +   \epsilon_{\rm L}\, W_{\rm L} \nonumber \\
& & \nonumber \\
& & + \sqrt{\epsilon_{\rm L}(1+  \epsilon)} W_{\rm TL} \cos\phi +  
\epsilon\, W_{\rm TT}\cos  2\phi\},
\label{eq:nonfact} 
\eea
where $\Gamma_{\rm V}$ is the flux of virtual photons, $\phi$ the out-of-plane
angle of the proton with respect to the electron scattering plane,
\eq 
\epsilon = 
\left [ 1 +   2{\vert{\vec q}\vert^2\over Q^2} \tan^2\oneh\theta
\right]^{-1},\qquad
\epsilon_{\rm\scriptstyle L} = {Q^2\over\vert{\vec q}\vert^2} \epsilon ,
\ee
and $Q^2 = \vert{\vec q}\vert^2 - \omega^2$ is the negative mass squared of the
virtual photon defined in terms of the momentum ${\vec q}$ and energy $\omega$
transferred by the incident electron through a scattering angle $\theta$. In
plane-wave impulse approximation, all structure functions become proportional to
the hole spectral functions and eq. (\ref{eq:factorised}) is recovered.

It turns out that for removal of valence protons a distorted-wave impulse
approximation (DWIA) is a suitable one~\cite{book96,rep93}. On the contrary, at
high missing energy and/or high missing momentum clear evidence for a better
approximation has been accumulated~\cite{ulmer87}-\cite{bob94}. In addition, other
processes beyond the simple one-body mechanism become important above the threshold
of two-nucleon emission and in the socalled dip-region, i.e. in the region between
the QF peak and the $\Delta$-resonance excitation~\cite{lourie86,kester95}.

In the dip region the semi-inclusive $^{12}$C\eep data of ref.~\cite{kester95}
have been compared with two calculations, one focusing on two-body meson-exchange
and $\Delta$ currents~\cite{jan94}, and the other one on short-range
correlations~\cite{ciofi91}. However, in this region many-body effects leading to
multi-nucleon emission are important. Limitations of the two-body process as a
mechanism for understanding the \eep reaction have been indicated in
refs.~\cite{tak89,gp94}.

In the QF region the one-body mechanism is dominant, but multiple
scattering of the ejected proton with the residual nucleus is also important as
shown by the large effects introduced by DWIA. A detailed analysis of the effects
of  multiple scattering has only been performed in a classical approach by means of
a Monte Carlo study~\cite{mc85} where an \eep reaction in a given nucleus is
simulated by taking into account multiple Coulomb and nuclear scattering by the
outgoing proton while crossing through the residual nucleus. A quantum-mechanical
treatment of FSI is proposed in the present paper taking advantage of the
multistep direct (MSD) scattering theory of Feshbach, Kerman and
Koonin~\cite{fkk80}.

The MSD theory has been extensively applied to describe the continuum spectrum in
nuclear reactions for energies up to the pion threshold (see~\cite{gad92} and
references therein;  \cite{chad93}-\cite{kon93}) establishing  the validity of the
theory over a wide range of energies and target nuclei. The reactions are described
as a series of two-body interactions between the projectile and the target nucleons
leading to the excitation of intermediate states of increasing complexity. At each
stage a nucleon may be emitted contributing to the pre-equilibrium energy spectrum.
The theory combines a quantum-mechanical treatment of multistep scattering with
statistical assumptions that lead to the convolution nature of the multistep
cross sections and enables the calculation of higher order contributions -- up to
six steps -- which would otherwise be impracticable.

In the present paper we apply the multistep scattering theory to describe the 
continuum spectrum of the QF \eep knockout reaction. Following the
electromagnetic interaction between the scattered electron and the target nucleus,
a target nucleon is excited to the continuum with energy $E_{1}$ and angle
$\Omega_{1}$ and subsequently undergoes a series of two-body interactions with the
residual nucleons before being emitted with energy $E$ and angle $\Omega$. We aim
to give a quantitative estimate of the multi-scattering effects in the high missing
energy region as a first step in the study of FSI.

In sect. 2 the MSD theory is briefly recalled and adapted to describe the proton
emission in \eep reactions. Calculational details are given in sect. 3 and
the results obtained in a case example are discussed in sect. 4.

\section{Theory}

The MSD theory has been described in detail in refs.~\cite{gad92,bon94} so here we
give  a brief account of the theoretical formalism adapted to \eep reactions 
without details of the derivations. The average cross section for an 
ejectile electron of energy $E_{k'}$ and angle $\Omega_{k'}$ and ejectile proton 
of energy $E$ and angle $\Omega$ is written as an incoherent sum of a one-step and
multistep ($n$-step) terms
\eq
\frac{{\rm d}^{4}\sigma}{{\rm d}\Omega_{k'} {\rm d}E_{k'} {\rm d}\Omega {\rm d}E} 
= 
\frac{{\rm d}^{4}\sigma^{(1)}}{ {\rm d}\Omega_{k'} {\rm d}E_{k'} {\rm d}\Omega
{\rm d}E } 
+ \sum_{n=2}^{\infty}\frac{{\rm d}^{4}\sigma^{(n)}}{ {\rm d}\Omega_{k'}
{\rm d}E_{k'} {\rm d}\Omega {\rm d}E},
\label{eq:msdeep}
\ee
where the $n$-step term is given by a convolution of the direct \eep
knockout cross sections and one-step MSD cross sections over all intermediate
energies $E_{1},\, E_{2}\dots$ and angles $\Omega_{1},\,\Omega_{2}\dots$ obeying
energy and momentum conservation rules:
 
\begin{eqnarray}
\frac{{\rm d}^{4}\sigma^{(n)}}{ 
{\rm d}\Omega_{k'}{\rm d}E_{k'} {\rm d}\Omega {\rm d}E} 
& =&  \left(\frac{m}{4\pi^{2}}\right)^{n-1} 
\int {\rm d}\Omega_{n-1}\int {\rm d}E_{n-1}E_{n-1}\dots \nonumber \\
& & \times \int {\rm d}\Omega_{1}\int {\rm d}E_{1}E_{1} 
\frac{{\rm d}^{2}\sigma^{(1)}}{{\rm d}\Omega dE}(E,\Omega
 \leftarrow E_{n-1},\Omega_{n-1})\dots \nonumber \\
& &\times \frac{{\rm d}^{2}\sigma^{(1)}}{{\rm d}\Omega_{2}
 {\rm d}E_{2}}(E_{2},\Omega_{2} \leftarrow E_{1},\Omega_{1}) 
\frac{{\rm d}^{4}\sigma}{{\rm d}\Omega_{k'}{\rm d}E_{k'} {\rm d}\Omega_{1} dE_{1}}.
\label{eq:msdneep}\\ \nonumber
\end{eqnarray}

\noindent
The cross section for the \eep direct knockout reaction is given by eq.
(\ref{eq:nonfact}) after having included the energy distribution (see eq.
(\ref{eq:distrib}) below). The one-step MSD cross sections for the subsequent NN
scatterings are calculated by extending the DWBA theory to the continuum and can be
written as \begin{eqnarray} & & \frac{{\rm d}^{2}\sigma^{(1)}}{{\rm d}\Omega {\rm
d}E} (E,\Omega \leftarrow E_{0},\Omega_{0}) \nonumber \\
& &\qquad\qquad =\sum_{J}(2J+1)\rho_{1{\rm p}1{\rm h},J}(U)
\left\langle\frac{{\rm d}\sigma(E,\Omega \leftarrow E_{0},\Omega_{0})}
{{\rm d}\Omega}\right\rangle^{\rm DWBA}_{J},
\label{eq:onestep}\\ \nonumber
\end{eqnarray} 
where $J$ is the orbital angular momentum transfer, $\langle {\rm d}\sigma/
{\rm d}\Omega\rangle^{\rm DWBA}_{J}$ is the average of DWBA cross sections exciting
1p1h states consistent with energy, angular momentum and parity conservation and
$\rho_{1{\rm p}1{\rm h},J}(U)$ is the density of such states with residual nucleus
energy $U=E_{0}-E$. The latter is factorised into a level-dependent density and a
spin distribution, $\rho_{1{\rm p}1{\rm h},J}(U) =  \rho_{1{\rm p}1{\rm
h}}(U)\,R_{n}(J)$. The energy-dependent density $\rho_{1{\rm p}1{\rm h}}$  is
obtained from an equidistant  Fermi-gas model with finite hole-depth restrictions
taken into account.  $R_{n}$ is a Gaussian spin distribution,
\begin{eqnarray}
R_{n}(J) = 
\frac{(2J+1)}{2(2\pi)^{1/2}\sigma_{n}^{3}}\exp\left[-\frac{(J+\oneh)^{2}}{2
\sigma_{n}^{2}}\right] ,
\end{eqnarray}
 with $\sigma_{n}$ the spin cut-off parameter. 
The transitions are
induced by an effective
NN interaction  which is given by a finite-range Yukawa potential with
strength $V_{0}$ adjusted to reproduce the experimental (p, p$'$) 
cross sections.

\section{Calculational details}
 
The QF \eep cross sections were calculated in DWIA~\cite{rep93}, including the
effect of Coulomb distortion of the electron waves, through the effective momentum
approximation, which is a good approximation for light nuclei~\cite{gp88}. A full
out-of-plane kinematics was considered, with an outgoing-proton energy up to the
maximum value compatible with the energy distribution of the bound single-particle
states.

The distorted waves were thus obtained from the optical potential of
ref.~\cite{gian76} which extends up to energies of 150 MeV and the bound-state
wavefunctions from a Woods-Saxon potential with a radius parameter $r_{0} = 1.3$
fm, diffuseness $a = 0.6$ fm~\cite{mou76} and a depth fixed to reproduce the input
energy eigenvalues. The quantum numbers and energy eigenvalues of the states that
can be excited were obtained from a spherical Nilsson shell model
scheme~\cite{seg67}. Such a scheme has been adopted because it is easily extended
into the continuum as required in calculating the MSD cross section. The price of 
consistency between bound and continuum states is, however, paid by removal
energies that are somehow different from experimental values.

The energy distribution of the bound states was taken as a
Lorentzian~\cite{mahaux88}

\eq
S(E_m) = \frac{2}{\pi}\frac{\Gamma(E_m)}
{4 (E_m - E_{\rm F} - E_{\rm b.e.})^2 - \Gamma^2(E_m)},
\label{eq:distrib}
\ee
where $E_{\rm b.e.}$ is the g.s. nucleon binding energy and $E_{\rm F}$ the 
Fermi energy of the Nilsson level scheme. The energy dependent width is given
by~\cite{br81}

\eq
\Gamma(E_m) = \frac{24(E_m - E_{\rm F})^2}{500 + (E_m-E_{\rm F})^2} ,
\label{eq:width}
\ee
where the energies are in MeV.

The MSD cross sections were calculated using DWUCK4~\cite{kun93} to obtain the
microscopic DWBA cross sections with a Yukawa effective NN potential  of range 1
fm. The strength of the potential $V_{0}$ was extracted from previous studies of
the systematics of the (p,p$'$) reaction on the atomic mass $A$ and  the incident
energy~\cite{ric94,chad94}. For sake of consistency the same distorted waves and
bound-state wavefunctions were used for the calculation of the \eep and DWBA
(p,p$'$) cross-sections. 

The microscopic transitions are averaged over transferred angular momentum and
residual nucleus energy by the MSD code~\cite{ola92} according to eq.
(\ref{eq:onestep}) where the 1p1h state density was calculated with an average
single-particle density $g=A/13$~\cite{gad92} and the spin  cut-off parameter of
its spin distribution was given by $\sigma^{2}_{2}  = 0.28\times 2\times
A^{2/3}$~\cite{fu86}. When calculating the multistep cross sections with eq.
(\ref{eq:msdneep}) the one-step MSD cross sections are obtained at several incident
energies lower than that of the excited proton of the direct \eep knockout
reaction and interpolated for other values. The convolution integral in eq.
(\ref{eq:msdneep}) is then evaluated using Monte Carlo integration (MSD
code~\cite{ola92}).

\section{Results}

The approach to MSD scattering of the ejectile proton presented in the previous
sections was applied to the $^{40}$Ca\eep reaction as a case example. The
$^{40}$Ca nucleus is an appropriate target for the statistical treatment of the
MSD theory and data at high missing energy exist with the following
kinematics~\cite{mou76}: the incident electron energy is $E_k =  497$
MeV,the electron scattering angle $\theta = 52.9^{\circ}$ and the outgoing proton
energy $E = 87\pm 10$ MeV. The scattered electron energy $E_{k'}$ varied in the
experiment from 350 to 410 MeV. In our calculations we fixed $E_{k'} =  350$ MeV
and worked at constant $({\vec q}, \omega)$ by varying the proton energy
$E$ accordingly. This kinematics is unable to reach missing momenta $p_m\simle 100$
MeV/$c$ for the deep states, contrary to the experimental situation, where the
detector acceptances also allow to probe low values of missing momenta.

In fig. 1 we show the theoretical direct \eep knockout and multistep cross
sections as a function of the angle $\gamma$ between the emitted proton ${\vec p}'$
and the momentum transfer ${\vec q}$ at four different residual nucleus energies
$U_{\rm res}= E_m - E_{\rm b.e.}$. At the lowest excitation energies, the direct nucleon knockout
process dominates and  exhibits a strong forward-peaking. The multistep
contributions are important at large  scattering angles over the whole energy
range. With increasing excitation energy the two-step and three-step contributions
become gradually more important than the one-step direct process over most of the
angular range, apart from the very small scattering angles $\gamma \leq
10^{\circ}$. The domination of multistep  processes at large scattering angles is
expected since as a result of multistep scattering the leading proton gradually
loses memory of its initial direction yielding thus increasingly symmetric angular
distributions. 

In order to compare with data it is useful to define the reduced cross section as
\eq
\rho(p_m,E_m) = \frac{1}{\sigma_{\rm cc1}}
{{\rm d}^3\sigma\over{\rm d}\Omega_{k'}{\rm d}E_{k'}{\rm d}\Omega} ,
\label{eq:reduced}
\ee
where $\sigma_{\rm cc1}$ is the electron-nucleon cross section taken according to
ref.~\cite{def83}.

In fig. 2 we compare $\rho(p_m,E_m)$ integrated over two different energy ranges
with the experimental data of ref.~\cite{mou76}. The theoretical curves are
multiplied by 0.5, a factor that can be interpreted as an average spectroscopic
factor. At low missing energies the direct process dominates, whereas at high
$E_m$ the major contribution comes from two-step and three-step processes. The
relative importance of multistep processes increases with the missing momentum and
is analogous to the behaviour of the multistep cross sections at large scattering
angles in fig. 1.

\section{Conclusions}

We have calculated the multi-nucleon-nucleon scattering contributions in \eep
reactions in the quasi-free region using the MSD theory of ref.~\cite{fkk80}, and
have shown that such processes are important at high missing energy and momentum.
Therefore, one can foresee that in
other kinematics involving even higher missing energy and momentum values like,
e.g., in the dip region~\cite{tak89}, multistep scattering processes would be
helpful in determining the final-sate interaction.


\bigskip

The authors are grateful to P.~E.~Hodgson for useful discussions.

\clearpage


\clearpage


\centerline{\bf Figure captions}

\medskip

Fig. 1. Differential cross section for the $^{40}$Ca\eep reaction as a function of
the angle $\gamma$ between the emitted proton and the momentum transfer at four
different energies $U_{\rm res}$ of the residual nucleus. Solid line for the direct
\eep process; dashed, dot-dashed and dotted lines for the two-, three- and four-step
processes. The total result is given by the solid line with marking dots.

\smallskip

Fig. 2. Reduced cross section for the $^{40}$Ca\eep reaction as a function of the
missing momentum integrated over two different missing-energy ranges. Experimental
data from ref.~\cite{mou76}, line convention for the theoretical curves as in 
fig. 1.


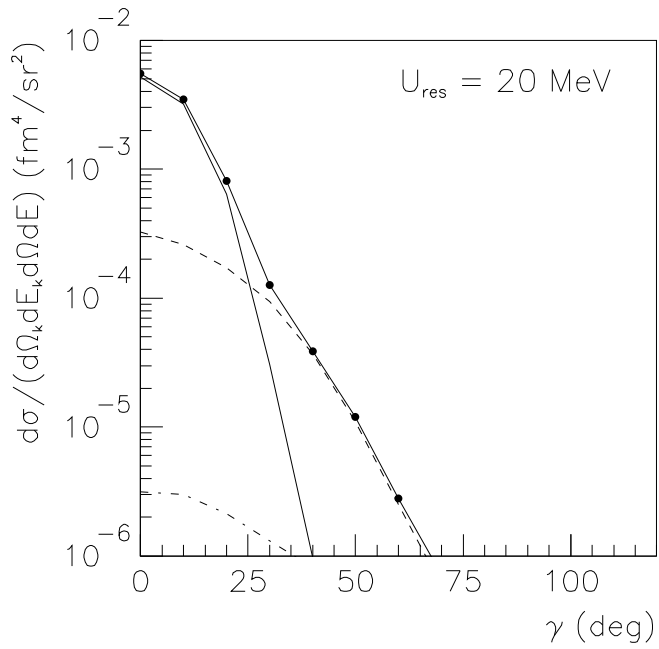
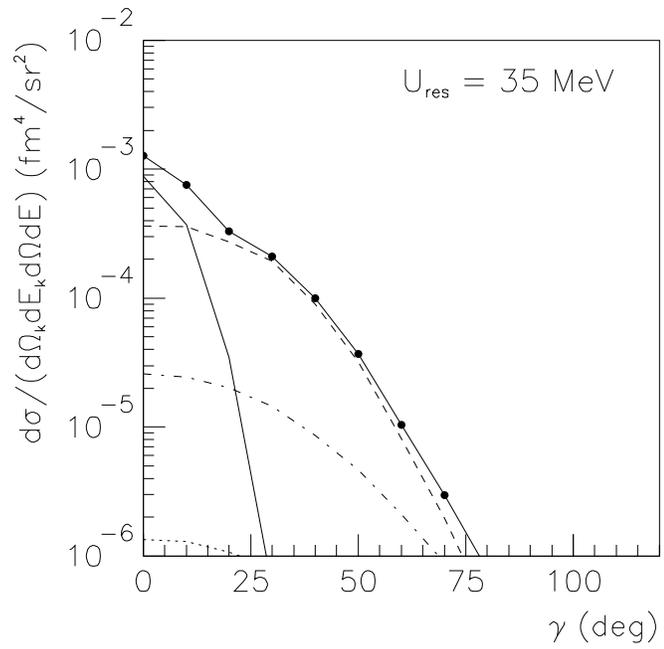
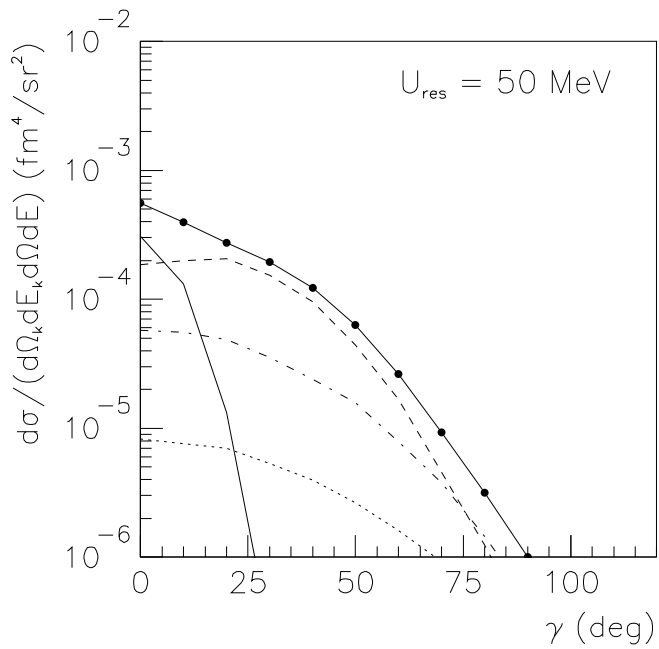
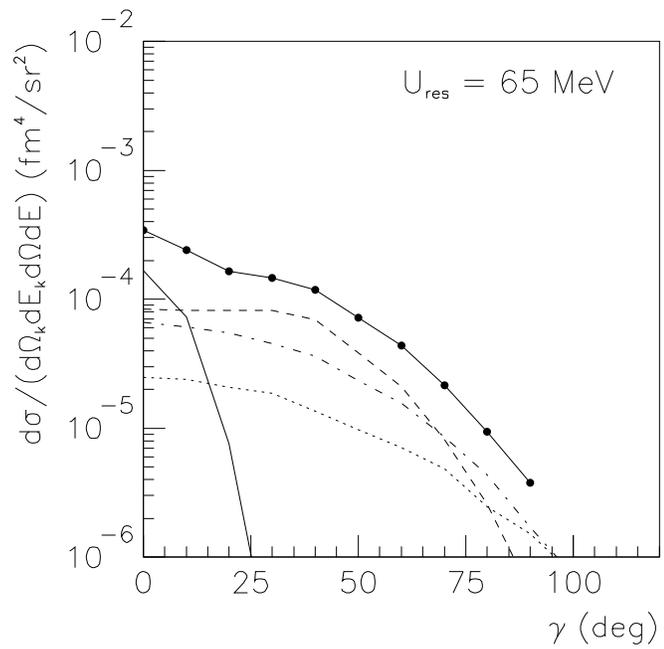
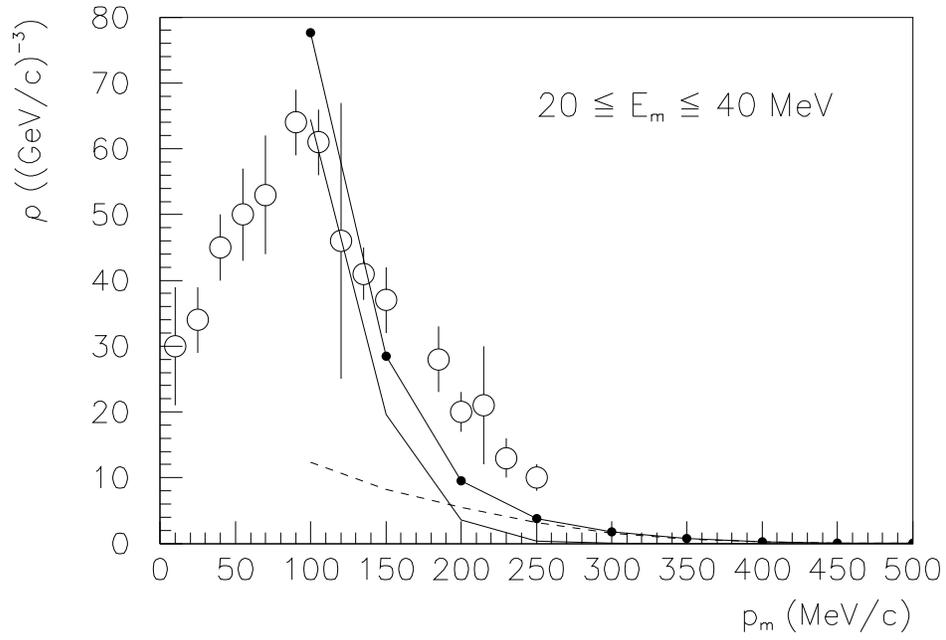
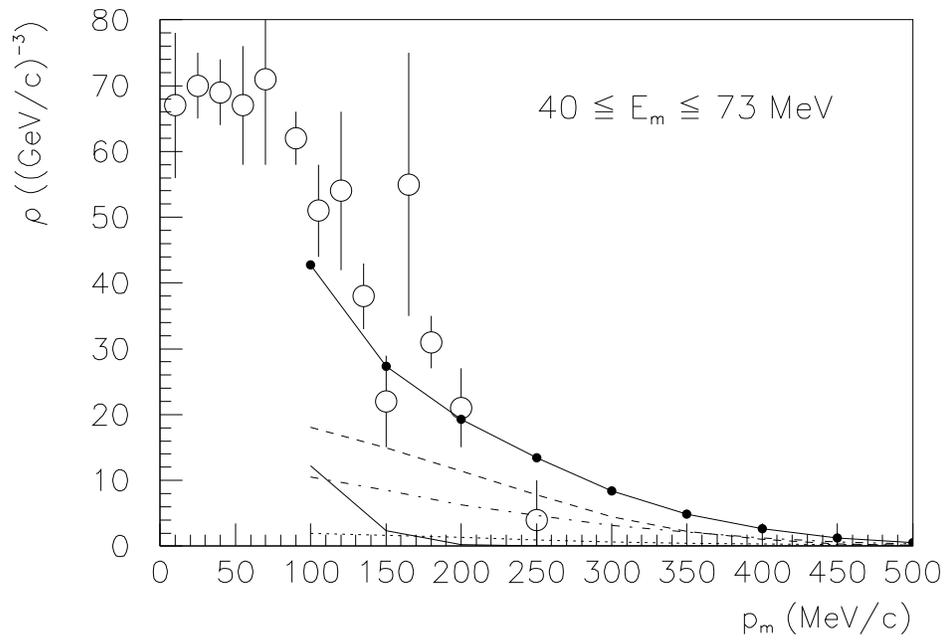
\end{document}